\documentstyle[preprint,prb,aps]{revtex}
\tightenlines
\title{A Hamiltonian flow associated with two dimensional map}
\author{Satoru SAITO\footnote{E-mail: saito@phys.metro-u.ac.jp\\
The work is supported in part by a Grant-in-Aid for general 
Scientific Research from the Ministry of Education, Sciences, 
Sports and Culture, Japan (No 10640278).
}, Akira SHUDO\footnote{E-mail: shudo@phys.metro-u.ac.jp},\\ 
Jun-ichi YAMAMOTO\footnote{E-mail: yjunichi@phys.metro-u.ac.jp} 
and Katsuhiko YOSHIDA$^\dagger$\footnote{E-mail: yoshida@kiso.phys.metro-u.ac.jp}}
\address{ Department of Physics, Tokyo Metropolitan University,\\
Minamiohsawa 1-1, Hachiohji, Tokyo 192-0397 Japan\\
$^\dagger$ School of Science, Kitasato University,\\ 1-15-1 
Kitasato
Sagamihara, Kanagawa, 228-8555 Japan
}
\begin{document}
\maketitle
\begin{abstract}
For a given differentiable map $(x,y)\rightarrow (X(x,y),Y(x,y))$, which has an inverse, we show that 
there exists a Hamiltonian flow in which $x$ plays the role of the time variable while $y$ is fixed.
\end{abstract}
\pacs{\#, 45.05.+x 45.20.Jj}

Let us consider a differentiable map 
\begin{equation}
(x, y)\ \rightarrow\ (X(x,y),Y(x,y)),
\label{(x, y)rightarrow (X,Y)}
\end{equation}
and assume that it has an inverse
\begin{equation}
(X,Y)\ \rightarrow\ (x, y).
\label{(X,Y)rightarrow (x, y)}
\end{equation}
A small change of $(x, y)$ causes a variation of $(X,Y)$. They are governed by
\begin{equation}
\left(\matrix{\partial_{x}\cr
\partial_{y}}\right)
=
J\left(\matrix{\partial_{X}\cr
\partial_{Y}}\right)
,\qquad 
J:=\left(\matrix{J_{xX}&J_{xY}\cr J_{yX}&J_{yY}}\right),
\label{1}
\end{equation}
where $\partial_x:=\partial/\partial x,\ J_{xX}:=\partial X/\partial x,\  {\rm etc}$,
or
\begin{equation}
\left(\matrix{\partial_{X}\cr
\partial_{Y}}\right)
=J^{-1}\left(\matrix{\partial_{x}\cr
\partial_{y}}\right).
\label{2}
\end{equation}

To be specific we consider the case in which 
$y$ is fixed while $x$ is changed. We introduce the notation
\[
Q(x):=X(x,y),\qquad P(x):=Y(x,y).
\]
Then from (\ref{1}) the variations of $(Q, P)$ are given by
\begin{equation}
{dQ\over dx}=J_{xX},\qquad {dP\over dx}=J_{xY}.
\label{3}
\end{equation}
We can prove the following: 

\vglue 0.5cm
\noindent
{\bf Proposition 1}

{\it Let $H$ be a function of $(Q, P)$ which is given by
\begin{equation}
H(Q,P)=\int^{y(Q,P)}(\det J)dy,
\label{Hamiltonian}
\end{equation}
and satisfies
\[
{\partial H\over\partial x}=0.
\]
Then the following system of Hamilton's equations hold:
\begin{equation}
{dQ\over dx}={\partial H\over\partial P},\qquad 
{dP\over dx}=- {\partial H\over\partial Q}.
\label{Hamilton's equations}
\end{equation}
Conversely, if $H(Q,P)$ satisfies Hamilton's equations, then it must be of the above form.
} 

Note that the value $x$ plays the role of time variable of this 
system. The proof is straightforward. Applying (\ref{2}) to $H$ yields
\begin{equation}
\left(\matrix{\partial_{Q}H\cr
\partial_{P}H}\right)
=
J^{-1}\left(\matrix{\partial_{x}H\cr
\partial_{y}H}\right).
\label{4}
\end{equation}
If we impose the condition that $H$ satisfies
\begin{equation}
J^{-1}\left(\matrix{\partial_{x} H \cr
\partial_{y} H}\right)
=
\left(\matrix{-J_{xY}\cr J_{xX}\cr}\right),
\label{7}
\end{equation}
and compare with (\ref{3}), the Hamilton equations (\ref{Hamilton's 
equations}) follow. To solve (\ref{7}) for $H$, we 
multiply $J$ from the left and obtain
\[
\left(\matrix{\partial_{x} H \cr
\partial_{y} H}\right)
=
J\left(\matrix{-J_{xY}\cr J_{xX}}\right)
=
\left(\matrix{0\cr
\displaystyle{{\partial X\over\partial x}}\displaystyle{{\partial 
Y\over\partial y}}
-
\displaystyle{{\partial X\over\partial y}}\displaystyle{{\partial 
Y\over\partial x}}\cr}\right),
\]
hence
\[
\partial_{x}H=0,\qquad 
\partial_{y}H={\partial X\over\partial x}{\partial Y\over\partial y}
-
{\partial X\over\partial y}{\partial Y\over\partial x}=\det J.
\]
Therefore (\ref{Hamiltonian}) is obtained. ({\it q.e.d.})

This paper is motivated by an observation found by 
one of the authors (A.S.) during the study of the quantized H\'enon 
map\cite{ShudoIkeda}, and generalizes the result to a wide 
class of maps. Before closing let us present the case of the H\'enon 
map, which is defined by
\begin{equation}
\left(\matrix{X\cr Y\cr}\right)
=
\left(\matrix{x^2-y+c\cr x\cr}\right),\qquad
\left(\matrix{x\cr y\cr}\right)
=
\left(\matrix{Y\cr Y^2-X+c\cr}\right),\qquad
J=\left(\matrix{2x&1\cr -1&0 \cr}\right).
\label{Henon}
\end{equation}
Since $\det J=1$, we have simply $H=y+ const$, hence
\begin{equation}
H(Q,P)=P^2-Q+c.
\end{equation}

\noindent
{\bf Acknowledgements}

One of the authors(A.S.) is grateful to Professor T. Aoki for his valuable
suggestions on exact WKB analysis. We also thank Professors M. Guest, Y.Ohnita and H.Isozaki for reading the manuscript and useful comments.


\end{document}